\documentclass[a4paper,11pt]{article}
\usepackage{jinstpub} % for details on the use of the package, please see the JINST-author-manual
\usepackage{subfig}
\usepackage{graphicx}
\title{\boldmath Model-Independent Machine Learning Approach for Nanometric Axial Localization and Tracking}

\author[a,1]{Andrey ALEXANDROV,\note{Corresponding author.}} % andrey.alexandrov@na.infn.it, {0000-0002-1813-1485}
\author[a,b]{Giovanni ACAMPORA,} % giovanni.acampora@unina.it, {0000-0003-4082-5616}
\author[a,b,c]{Giovanni {DE LELLIS},} % giovanni.delellis@unina.it, {0000-0001-5862-1174}
\author[a,b]{Antonia {DI CRESCENZO},} % antonia.dicrescenzo@unina.it,{0000-0003-4276-8512}
\author[b]{Chiara ERRICO,} % chiara261110@gmail.com
\author[d]{Daria MOROZOVA,} % dashia110999@mail.ru, {0009-0000-5650-5030}
\author[a]{Valeri TIOUKOV} % valeri.tioukov@na.infn.it, {0000-0001-5981-5296}
\author[a,b]{and Autilia VITIELLO} % autilia.vitiello@unina.it, {0000-0001-5562-9226}

\affiliation[a]{Istituto Nazionale di Fisica Nucleare (INFN) sezione di Napoli, Napoli 80126, Italy}
\affiliation[b]{Università degli Studi di Napoli Federico II, Napoli 80126, Italy}
\affiliation[c]{European Organization for Nuclear Research (CERN), Geneva, Switzerland}
\affiliation[d]{National University of Science and Technology MISIS, Moscow 119049, Russia}

\emailAdd{andrey.alexandrov@na.infn.it}

\abstract{Accurately tracking particles and determining their coordinate along the optical axis is a major challenge in optical microscopy, especially when extremely high precision is needed. In this study, we introduce a deep learning approach using convolutional neural networks (CNNs) that can determine axial coordinates from dual-focal-plane images without relying on predefined models. Our method achieves an axial localization precision of 40 nanometers—six times better than traditional single-focal-plane techniques. The model’s simple design and strong performance make it suitable for a wide range of uses, including dark matter detection, proton therapy for cancer, and radiation protection in space. It also shows promise in fields like biological imaging, materials science, and environmental monitoring. This work highlights how machine learning can turn complex image data into reliable, precise information, offering a flexible and powerful tool for many scientific applications.}

\keywords{Image processing; Data analysis; Particle tracking detectors; Dark Matter detectors (WIMPs, axions, etc.); Instrumentation for hadron therapy}

\arxivnumber{2505.14754} % Only if you have one

\begin{document}
\maketitle
\flushbottom

\section{Introduction}

Recent advancements in machine learning have significantly enhanced the precision and efficiency of data-driven methodologies in scientific applications. These methods have found applications in a variety of fields, including physics, medicine, and space sciences, where they help addressing complex challenges which require high-precision measurements. 

One such application is directional dark matter search experiments that require precise measurements of ions recoiling after their interactions with dark matter particles~\cite{DirectDM1,UFN_rev}. 
Due to their extremely low kinetic energies, in the $1 - 100$ keV range, recoiling ions produce tracks ranging from a few millimeters in gases at low pressure to a few hundreds of nanometers in solids~\cite{UFN_rev,Battat2016}. Taking into account that the required detector mass in practice amounts to several tons, the choice of solid materials as a sensitive medium is advantageous.

The only detector technology that has shown so far to be capable of reconstructing sub-micrometer recoil tracks in a solid medium is the nuclear emulsion~\cite{NEWSdm:2017efa}. It has been demonstrated that it is possible to reconstruct tracks with lengths down to 50~nm using special emulsion films with nanometric crystals and a custom-designed super-resolution optical microscope~\cite{alexandrov2023super}. In general, super-resolution refers to any imaging technique (optical, computational, or hybrid) that achieves spatial resolution beyond the classical diffraction limit of light, typically around 200–250~nm in the transverse plane and 500–700~nm in the axial direction for visible wavelengths. In the context of the study reported in~\cite{alexandrov2023super}, super-resolution is achieved in the transverse plane, where the system reaches resolutions of approximately 50–100 nm. However, the axial resolution remains diffraction-limited. Enabling the super-resolution also in the axial direction will make it possible to achieve a precise 3D measurement of recoiling ion tracks, thus improving the overall detector sensitivity to dark matter by almost one order of magnitude~\cite{OHare2015}.

%It has been demonstrated that it is possible to reconstruct tracks with lengths down to 50~nm using special emulsion films with nanometric crystals and a custom-designed super-resolution optical microscope~\cite{alexandrov2023super}. However, the super-resolution is achievable only in the transverse plane while in the axial direction the resolution is still diffraction-limited. Enabling the super-resolution also in the axial direction will make it possible to achieve a precise 3D measurement of recoiling ion tracks, thus improving the overall detector sensitivity to dark matter by almost one order of magnitude~\cite{OHare2015}.

A similar challenge is present in proton therapy cancer treatment, where the fragmentation of nuclei in biological tissues produces short-ranged, highly ionizing particles~\cite{FOOT1}. The contribution of these fragments to relative biological effectiveness (RBE) is crucial for efficient treatment planning~\cite{RBE_ref}. However, due to their short range, precise measurements are difficult and often obtained using inverse kinematic approaches, which introduce systematic uncertainties~\cite{FOOT1,FOOT2}. Accurate 3D measurement of these fragments would allow direct measurement of fragmentation cross-sections, contributing to the benchmarking of Monte Carlo (MC) transport codes and improving the accuracy of treatments.

The fragmentation of light ions on light targets also plays a significant role in radioprotection in space, where secondary fragments produced by nuclear interactions with spacecraft components contribute to the dose delivered to astronauts and damage to electronic systems~\cite{Durante2011,Durante2014}. Accurate measurement of these fragments is essential for improving MC-based shielding designs.

One possible way to overcome the diffraction-limited axial localization and axial resolution problems is to use the dual-focal-plane imaging technique that simultaneously captures two separate focal planes within the specimen, providing extra information~\cite{MUM2}. However, this method relies on complex mathematical models to describe light interactions and propagation, which often involve numerous parameters that are difficult to determine precisely. 

This study introduces a model-independent, deep learning-based approach for nanometric axial localization and particle tracking using convolutional neural networks (CNNs) applied to dual-focal-plane images. Traditional techniques often rely on detailed system parameters and are hard to adapt to different imaging setups, limiting their use in real-time or cross-disciplinary applications. The novelty lies in that our model learns the relationship between image features and axial coordinates directly from image data. This eliminates the need for detailed knowledge of the optical system or the physical properties of the objects under study. Moreover, the model-independent nature of this technique makes it highly adaptable, allowing it to be applied to various imaging systems without requiring significant adjustments. This feature makes the method particularly valuable for multidisciplinary applications where the underlying imaging conditions or object properties may vary significantly.

%%%%%%%%%%%%%%%%%%%%%%%%%%%%%%%%%%%%%%%%%%

\section{Materials and Methods}

\subsection{Image dataset}
\label{sec:image_data}

% 0.6 mm^2
% 20 um thick
The images for this study were taken with a high-resolution, single-focal-plane optical microscope described in the Appendix section. The scanned sample contained silver nanoparticles with a diameter of 60~nm randomly immersed in gelatin. The whole thickness of 20~$\mu$m was sampled with steps in the range 245-255~nm, a value chosen independently of this study, based on prior research~\cite{alexandrov2020super} demonstrating that it provides a balance between axial resolution and data acquisition speed. Images of nanoparticles could appear anywhere in the field of view and were not necessarily in focus. Then the nanoparticle images, both focused and unfocused, were detected by a program, cropped, and saved to files along with the corresponding X, Y, and Z coordinates.

\begin{figure*}%[ht]
%\centering
\begin{center}
\subfloat[][\label{fig:1} $\Delta{z} = -540$~nm]
{\includegraphics[ width=.3\textwidth, height=
0.2\textheight]{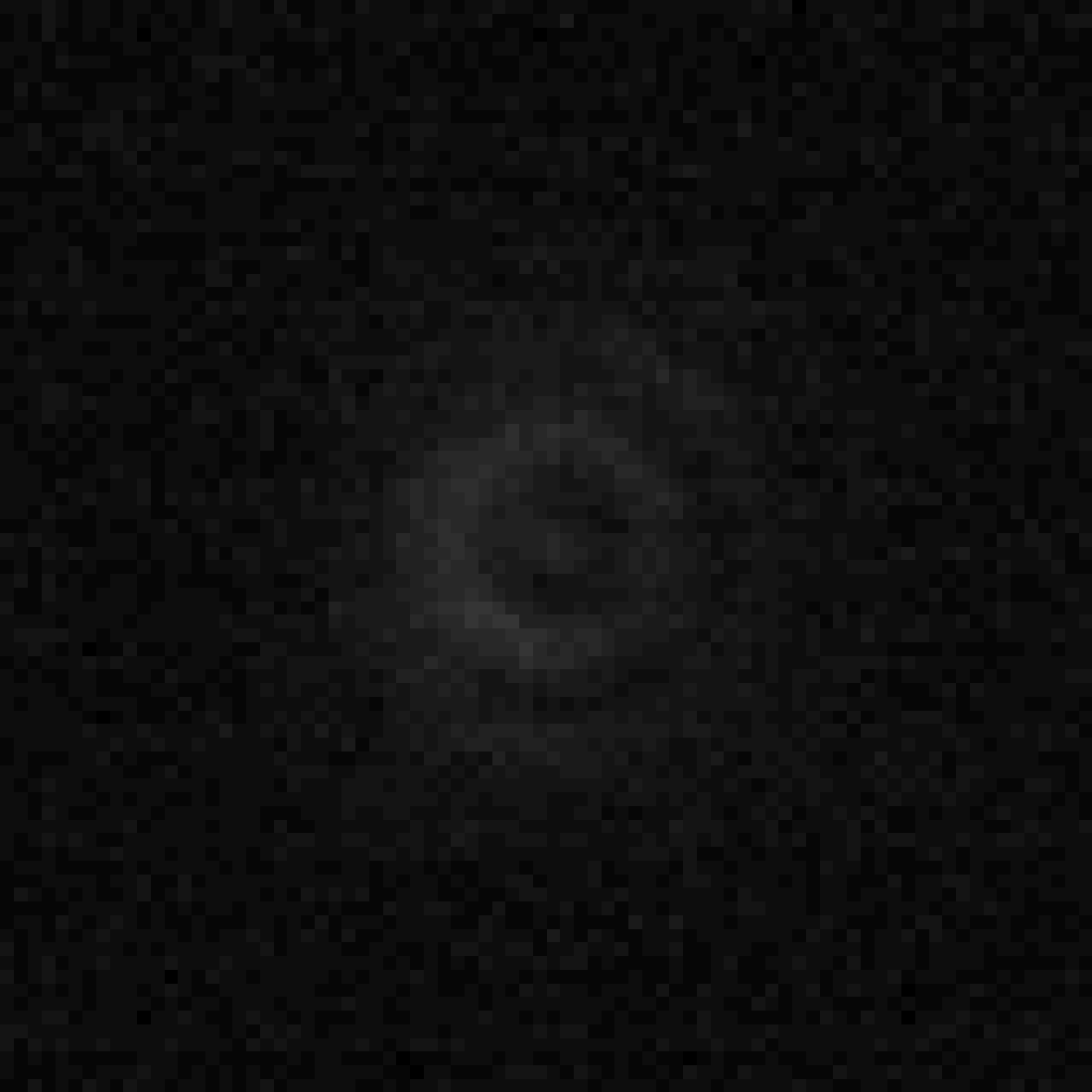}}\quad
\subfloat[][\label{fig:2} $\Delta{z} = -295$~nm]
{\includegraphics[ width=.3\textwidth, height=
0.2\textheight ]{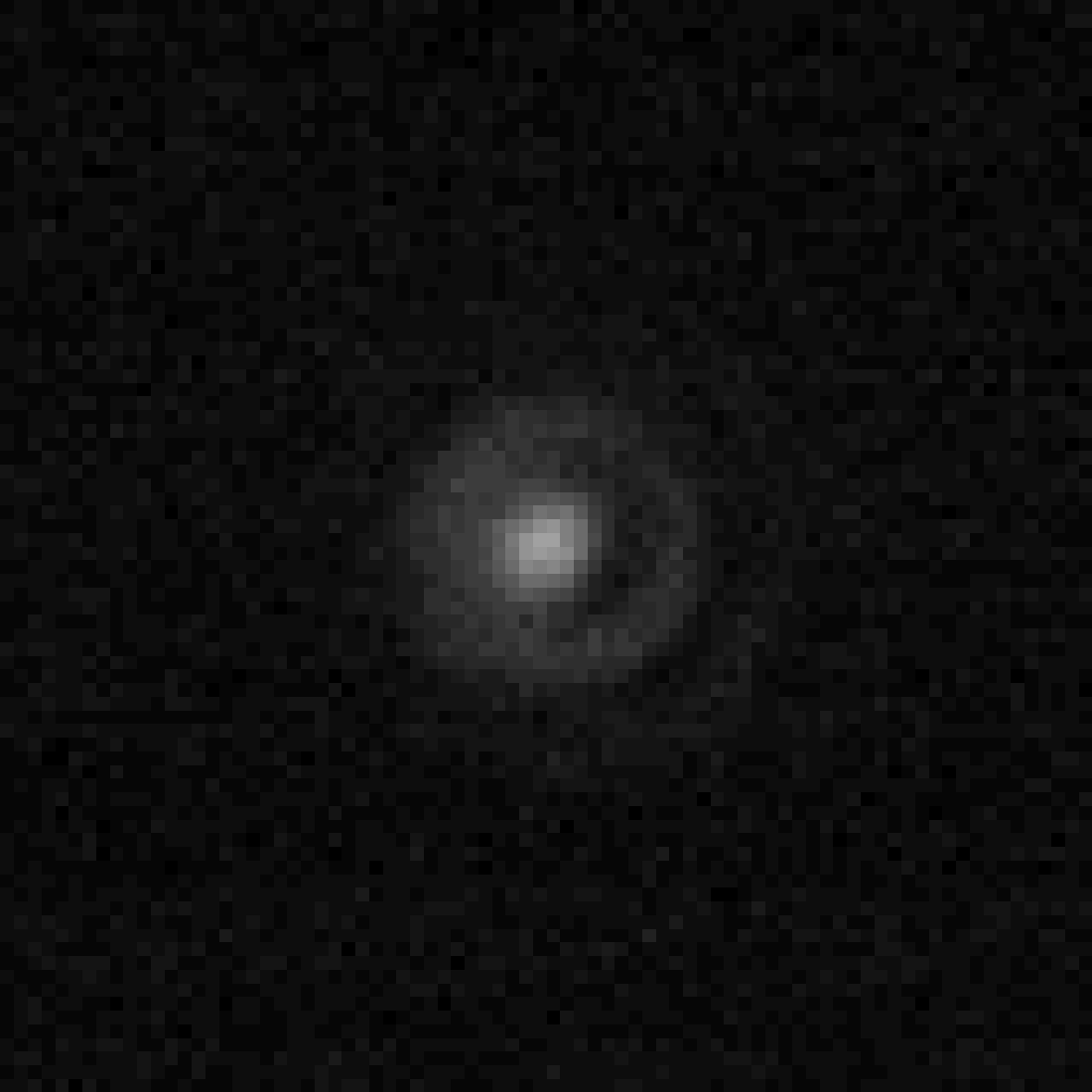}}\quad
\subfloat[][\label{fig:3} $\Delta{z} = -40$~nm]
{\includegraphics[width=.3\textwidth, height=
0.2\textheight]{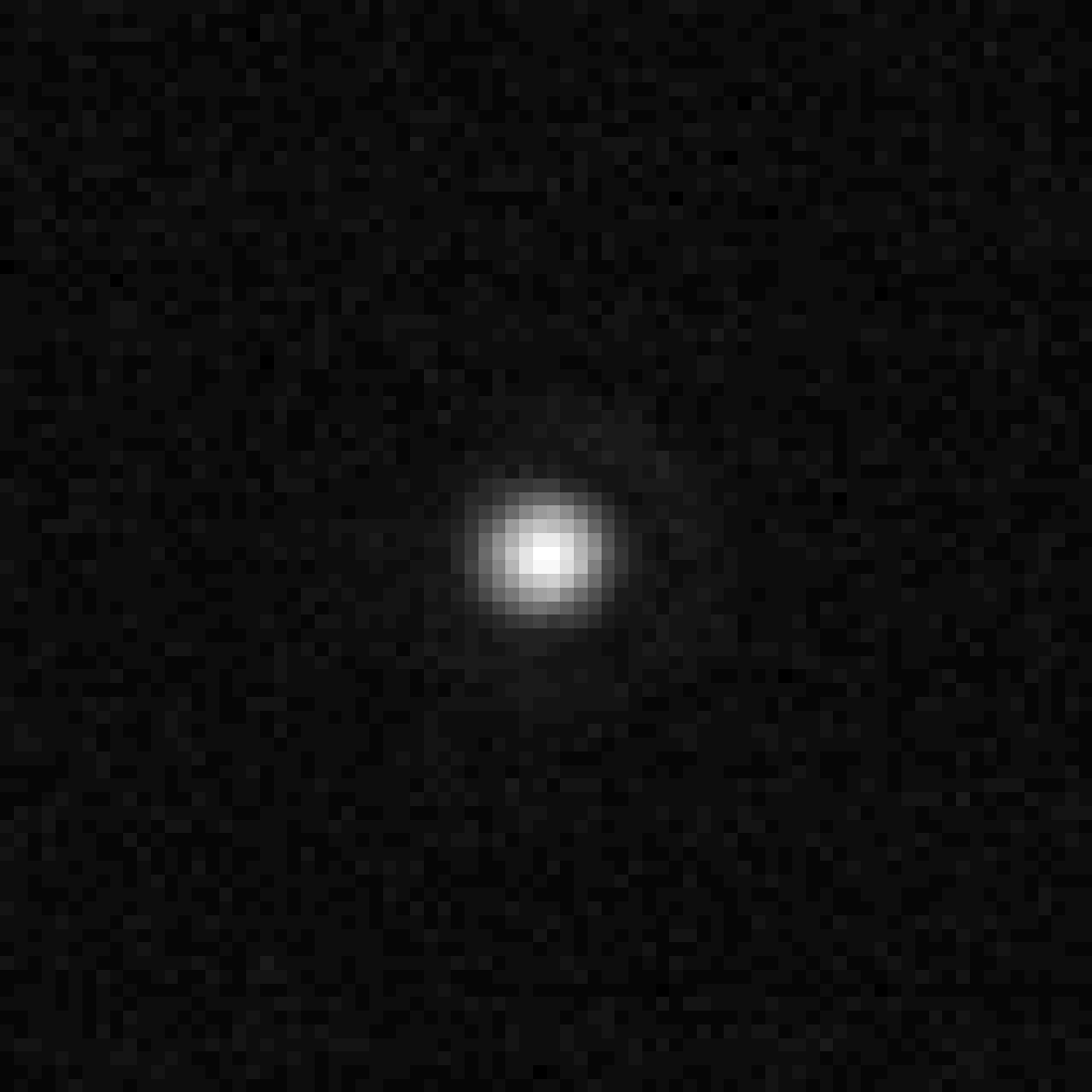}}\\
\subfloat[][\label{fig:4} $\Delta{z} = 205$~nm]
{\includegraphics[ width=.3\textwidth, height=
0.2\textheight]{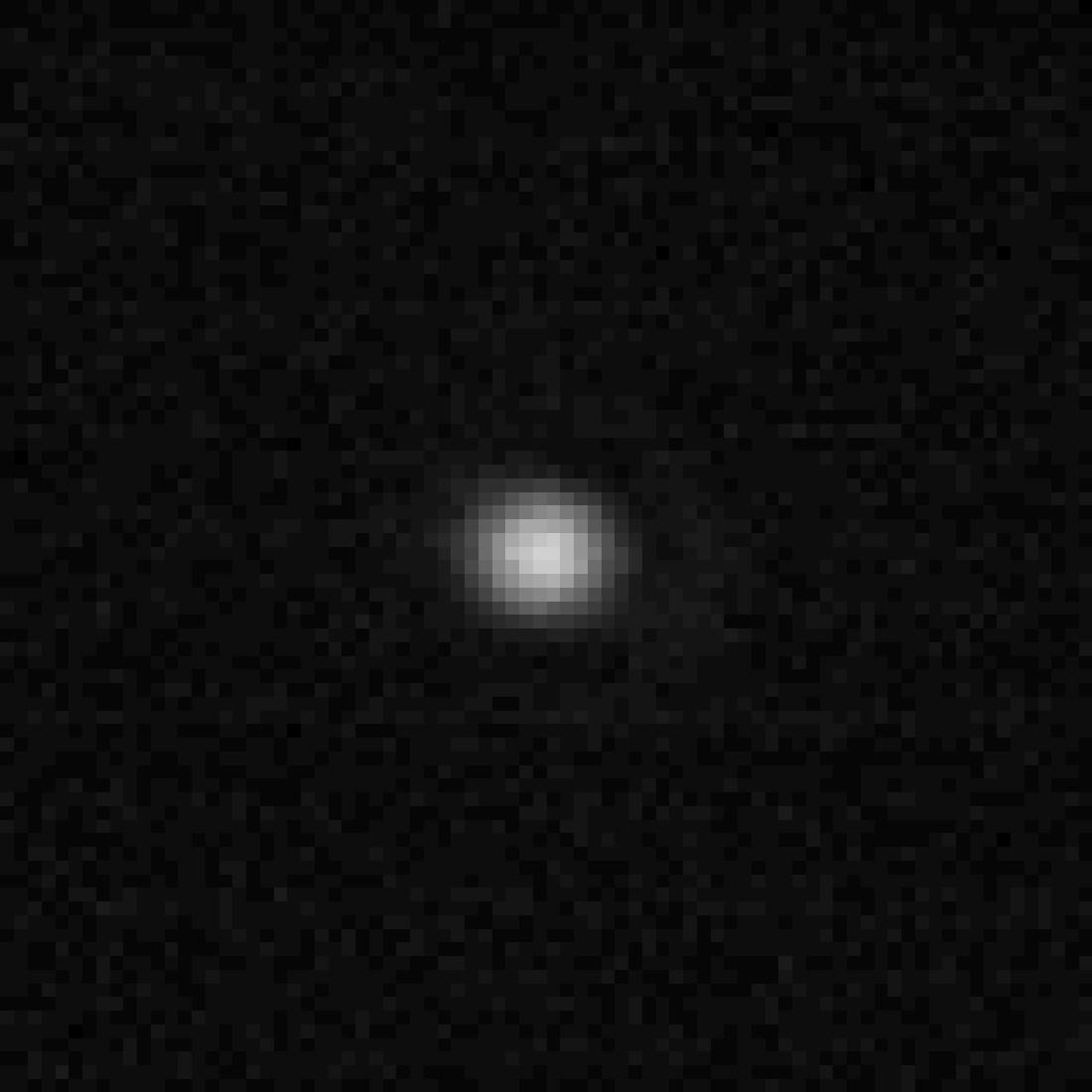}}\quad
\subfloat[][\label{fig:5} $\Delta{z} = 460$~nm]
{\includegraphics[ width=.3\textwidth, height=
0.2\textheight ]{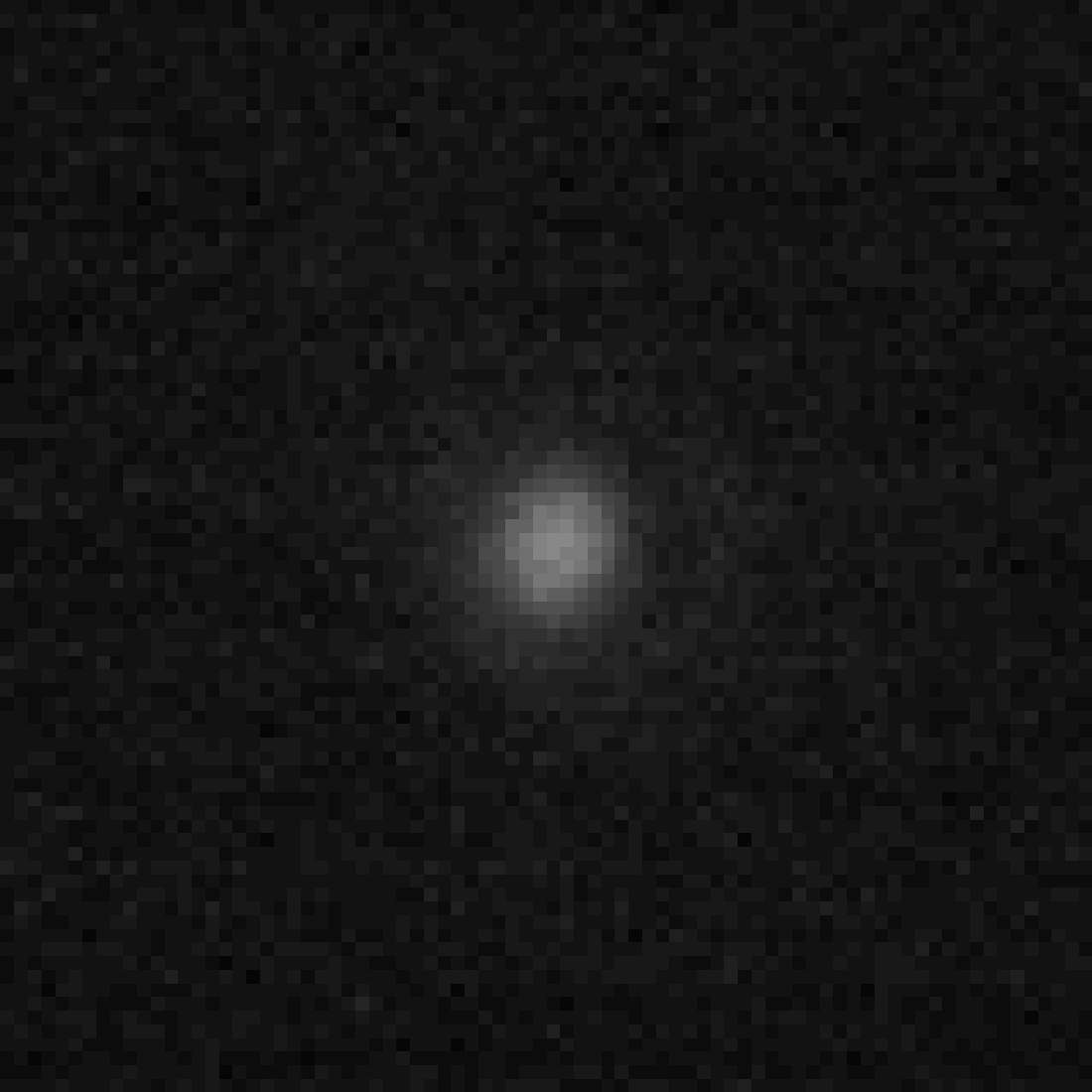}}\quad
\subfloat[][\label{fig:6} $\Delta{z} = 705$~nm]
{\includegraphics[width=.3\textwidth, height=
0.2\textheight]{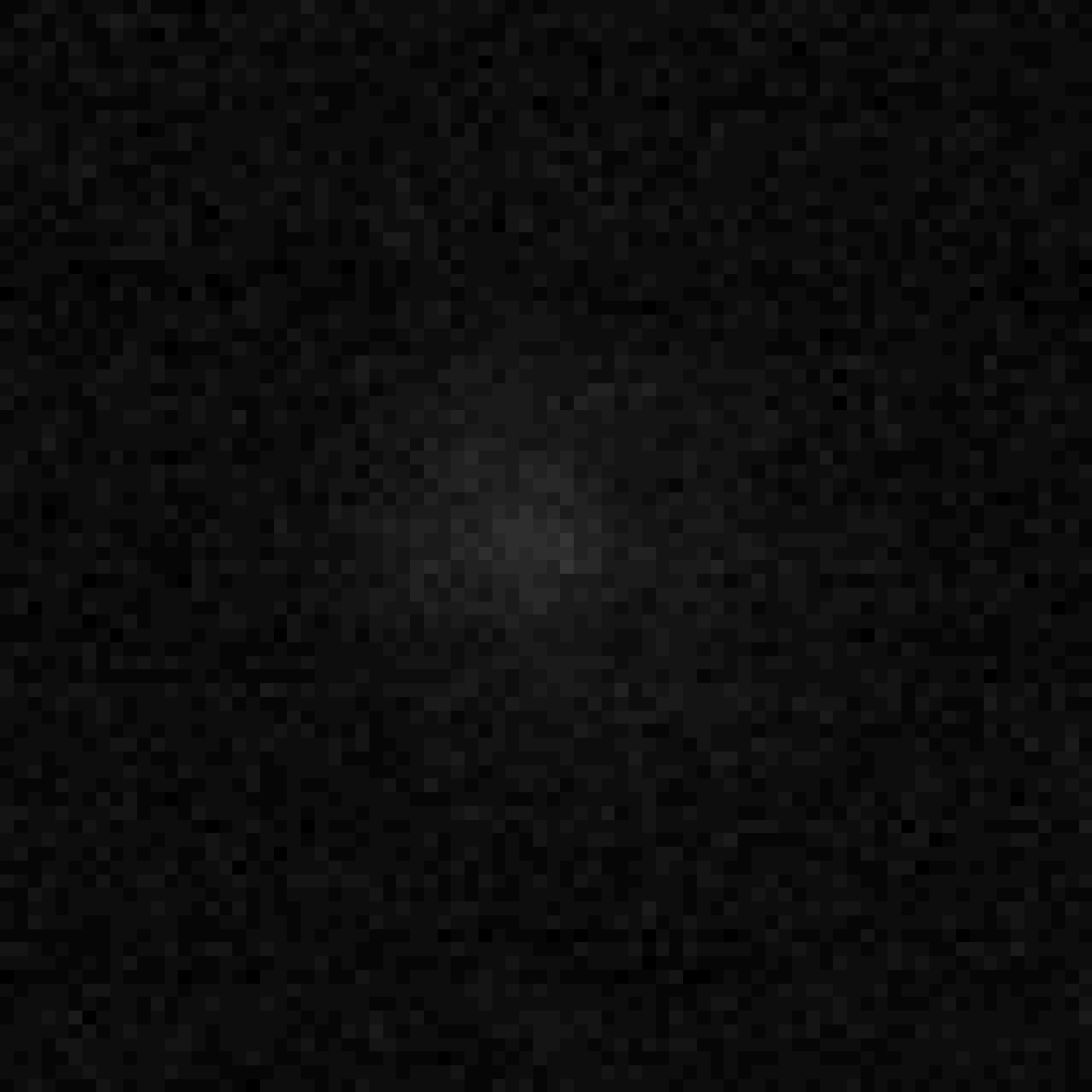}}
\caption{Images of the same nanoparticle acquired at different axial coordinates with step in the range $245-255$~nm. The dimensions of the image are $2150 \times 2150$~nm. The $Z$ axis is directed towards the objective lens.}
\label{fig:im}
\end{center}
\end{figure*}

\begin{figure}
\begin{center}
\includegraphics[width=.5\textwidth]{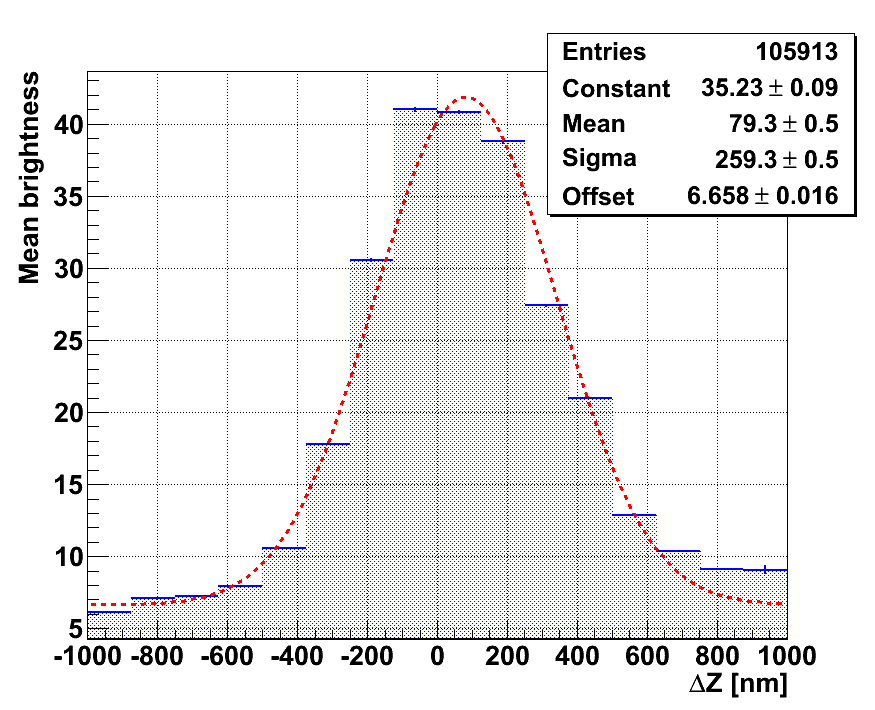}
\caption{Mean brightness profile along the optical axis averaged over 105913 nanoparticles. Error bar values are calculated as the RMS of the nanoparticle brightness distribution in the corresponding bin. The $Z$ axis is directed towards the objective lens.}
\label{fig:Z_prof}
\end{center}
\end{figure}

Figure~\ref{fig:im} shows an example of images of a single nanoparticle taken at different defocus values $\Delta{z}$. As a result of minor uncorrected spherical aberration, the Airy pattern is clearly visible predominantly in the negative $Z$ direction. In contrast, in the positive $Z$ direction, approaching the objective lens, the rings appear blurred. This asymmetry also causes the shape of the brightness profile of the nanoparticles, depicted in Figure~\ref{fig:Z_prof}, to differ slightly from a Gaussian distribution shown as a red dotted line. However, our deep learning approach is data-driven and does not assume a specific model, allowing it to adapt to these variations and maintain high localization precision.

%\begin{figure*}[ht]
%\end{figure*}

\subsection{Nanoparticle reference axial coordinate definition}
\label{sec:z_calc}

The sampling step in the axial direction of the image dataset used for this study is in the range $245-255$~nm, ensuring that a nanoparticle appears in at least three consecutive images. This allowed for the calculation of the axial coordinate using the weighted average of the brightness distribution:
\begin{equation}
    \label{eq:1}
   \bar{z_{i}} = \frac{\sum_{j}(b_{i,j} z_{j})}{\sum_{j}b_{i,j}},
\end{equation}
where $b_{i,j}$ is the average brightness of nanoparticle $i$ in image $j$ with axial coordinate $z_j$.

The axial localization precision is defined by:  
\begin{equation}
    \label{eq:2}
   \sigma_{\bar{z_{i}}}^{2} = \frac{\sum_{j}b_{i,j}^2}{(\sum_{j}b_{i,j})^2}(\sigma_{z}^{im})^2 + \frac{\sum_{j}(z_{j}-\bar{z_{i}})^2}{(\sum_{j}b_{i,j})^2}(\sigma_{B}^{im})^2,
\end{equation}
where $\sigma_{z}^{im} = 35$~ nm is the objective lens coordinate measurement precision and $\sigma_{B}^{im} = 0.73$ is the noise at the videocamera sensor, calculated as the RMS pixel brightness variation. The mean localization precision determined by the brightness profile using equation \ref{eq:2} is $\sigma_{z}^{prof} \equiv \bar{\sigma}_{\bar{z_i}} = (18 \pm 1)$~nm averaged over 105913 nanoparticles.

The value $\bar{z_{i}}$ calculated using equation~\ref{eq:1} is used in training as the reference axial coordinate of nanopaticle ${i}$. Each calculation involves several (typically 5-6) consecutive images taken at different axial coordinates by a single-focal-plane microscope. While this approach is effective for measuring static objects, it is not practical for real-time applications. Capturing multiple images sequentially requires time for stage movement and stabilization, making it too slow for dynamic objects. Consequently, a single-focal-plane microscope cannot benefit from dual-focal-plane image fitting, and its axial localization precision remains limited by the width of the brightness profile shown in Figure~\ref{fig:Z_prof}.

\subsection{Training datasets}
\label{sec:train_data}

This study aims to develop a novel measurement approach tailored for dual-focal-plane microscopes. To replicate two focal planes, we selected images from the existing dataset to form image pairs that are offset by $\delta$ relative to each other. For each pair of images, the defocus distance from the corresponding nanoparticle to the bottom image (with the smaller $Z$ coordinate) in the pair is calculated as $\Delta{z}_{k}=z_{k}^{bottom} - \bar{z}_{i}$, where $z_{k}^{bottom}$ is the axial coordinate of the bottom image in the pair $k$ and $\bar{z}_{i}$ is the axial coordinate of the nanoparticle $i$ to which the image pair $k$ belongs. By applying this procedure to all nanoparticles, we establish a set of image pairs $k$ along with the corresponding defocus distance $\Delta{z}_{k}$. Such a triplet: two images offset by $\delta$ and the defocus distance $\Delta{z}_{k}$, form an independent measurement carried out by a dual-focal-plane microscope.

We denote as ${\Delta{z}_{k}}^{test}$ the defocus distances calculated using $\bar{z}_{i}$ measured with the brightness profile and as ${\Delta{z}_{k}}^{pred}$ those predicted by CNN. Values of ${\Delta{z}_{k}}^{test}$ are used for training. During the test phase, CNN predicts ${\Delta{z}_{k}}^{pred}$ that are later compared with $\Delta{z}_{k}^{test}$ to estimate the localization precision.

In order to check the performance of the method at different offsets $\delta$, we prepared three datasets \cite{datasets} with $\delta$ equal to 250~nm, 500~nm and 750~nm. The total size of each image dataset is shown in the second column of Table~\ref{tab:1}.

\subsection{CNN model architecture and training}
\label{sec:cnn}

A convolutional neural network (CNN) is a deep learning model for processing data that has a grid pattern, such as images, which is inspired by the organization of animal visual cortex and designed to automatically learn spatial hierarchies of features \cite{yamashita2018convolutional}. CNNs are mainly supervised learning models where the input to the system and the desired output are known during training procedure. The goal of CNN is to learn the mapping between the input, typically an image, and its corresponding output by detecting a series of abstract representations of the characteristics, ranging from simple to more complex ones. In our study, the input is a combination of two images associated by considering a certain distance $d$ and the output is the real value $z$.

Similarly to traditional neural networks, CNNs are composed of several basic building blocks, called \emph{layers}. These layers are typically categorized in three types: convolutional, pooling and fully connected layers. The main peculiarity of CNNs that makes them different from traditional neural networks is represented by the first two layers: convolution and pooling. The convolutional layer is used to extract features from images through the use of filters which are convolved with a given input to generate an output feature map. A filter is used to detect the presence of specific features or patterns present in the original image (input). The training of a CNN model will consist of identifying the filter that works best on a given training dataset. To achieve this goal, an optimization algorithm uses a loss function that compares CNN predictions to known outputs to measure how well CNN predictions match what was expected. As for the pooling layer, it is aimed at enhancing the power of convolutions by reducing the size of the output feature map but, at the same time, still keeping and intensifying the useful features the filters show when convolving the image. Finally, the third type of layer, the fully connected layer, is used to map the extracted features into final output that can be categorical in the case of classification or a real value in the case of regression problems.

\begin{figure*}%[ht]
\begin{center}
\includegraphics[width=\textwidth]{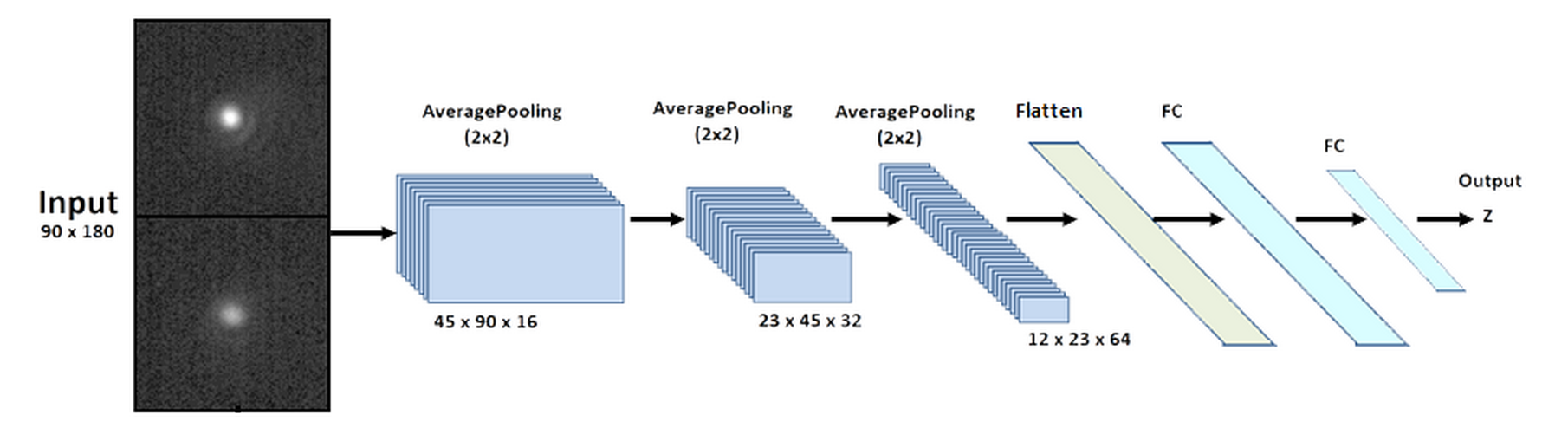}
\caption{Schematic representation of CNN architecture.\label{fig:Our_CNN}}
\end{center}
\end{figure*}

As reported in Figure \ref{fig:Our_CNN}, the implemented CNN for our study is composed of three convolutional layers with size 16, 32 and 64, respectively. The convolution operation is performed with a filter of $3\times3$ by using as padding the same strategy. The output of the convolution operation is passed through a nonlinear activation function. In this study, the most common nonlinear activation function, namely Rectified Linear Unit (ReLU), is used. This function returns either the same value given in input, or zero if the input value is negative. Formally,
\begin{equation}
    f_{relu}(x)=max(0,x)
\end{equation}
The output of the activation function is undergone to a Batch Normalization (BN) layer that applies a transformation that maintains the mean output close to 0 and the output standard deviation close to 1. Finally, the pooling layer downsamples the input representation by taking, in our case, the average value over the window defined by pool size ($2\times2$) for each dimension along the features axis. In order to provide input data to the two fully connected layers, a flatten layer is necessary to rearrange the three dimensional data, obtained in output from the last pooling layer, in a one dimensional vector. In particular, the last fully connected layer is added to perform regression. Therefore, the output of the first fully connected layer is passed through a ReLU activation function and a BN layer, whereas, the output of the second fully connected layer is passed through a linear activation function to perform regression. Moreover, a dropout layer is added in order to randomly set input units to 0 (removing neurons) with a frequency of rate that is equal to 0.5 between the two fully connected layer in order to prevent overfitting.

As for the training procedure, the CNN is trained by using Adam optimizer \cite{kingma2014adam}, that is an algorithm for first-order gradient-based optimization of stochastic objective functions, based on adaptive estimates of lower-order moments. The used loss function is the mean squared error metric that is commonly applied in the regression problem \cite{yamashita2018convolutional}. This metric computes the mean of squares of errors between the true $z$ value, denoted as $z$, and the predicted $z$ value, denoted as $\hat{z}$ as follows:
\begin{equation}
f_{loss} = \frac {1}{n} \sum _{i=1}^{n}(z_i - \hat {z_{i}})^{2}
\end{equation}
where $n$ is the number of instances contained in the training dataset. The training procedure involved a learning rate set to 0.001, a batch size set to 16, a number of epochs set to 400. The original image dataset was divided in training and test data with a percentage 80\%-20\%, respectively. 

The training parameters were selected based on a combination of standard practices in deep learning and a dedicated hyperparameter optimization study.
%conducted in a related thesis work~\cite{Errico:2019thesis}. In that study, the same CNN architecture was applied to the problem of axial localization using dual-focal-plane images of 60~nm silver nanoparticles. 
A systematic scan of hyperparameters was performed, including batch size (4 to 64), dropout rate (0.4 to 0.6), and number of epochs (30 to 600), while monitoring the resulting localization precision. The batch size of 16 consistently yielded the best trade-off between convergence stability and generalization, while 400 epochs provided sufficient training without overfitting. The learning rate of 0.001 was selected based on standard practice for the Adam optimizer and was found to be effective without requiring further tuning. 
%These parameters were adopted in the present study to ensure consistency and reproducibility. The results reported here (e.g., axial localization precision of 40~nm for $\delta = 500$~nm) confirm the robustness of this configuration across datasets and imaging conditions.

\section{Results}

\subsection{Deep Learning approach}
\label{sec:DL_arch}
To achieve high-precision axial localization, we implemented a CNN-based regression model designed to predict axial coordinates from dual-focal-plane image pairs. The network architecture consists of:
\begin{itemize}
\item Three convolutional layers with ReLU activation functions and batch normalization.
\item Average pooling layers to reduce spatial dimensions and extract hierarchical features.
\item Two fully connected layers with dropout regularization to prevent overfitting.
\item An output layer optimized for regression using the mean squared error loss function.
\end{itemize}

The CNN model transforms raw image pairs into axial coordinate predictions by learning implicit mappings between pixel intensity patterns and object axial coordinate. The extracted axial coordinates represent actionable knowledge, enabling precise 3D localization without explicit physical models. This data-driven approach offers a generalizable solution adaptable to diverse imaging systems and scientific applications.

\begin{table}[htbp]
\centering
\caption{Training datasets and performance.\label{tab:1}}
\smallskip
\begin{tabular}{c c c c}
\hline
\textbf{Offset $\delta$, nm} & \textbf{Image pairs} & \multicolumn{2}{c}{\textbf{Localization precision, nm}} \\
%\cline{3-4}
 & & \textbf{Test} & \textbf{Training} \\
\hline
250 & 33562 & $48.4 \pm 0.7$ & $46.2 \pm 0.6$ \\ 
500 & 34842 & $40.0 \pm 0.5$ & $38.9 \pm 0.5$ \\ 
750 & 21685 & $44.0 \pm 0.7$ & $42.6 \pm 0.6$ \\ 
\hline
\end{tabular}
\end{table}

%\begin{table}[htbp]
%\centering
%\caption{Training datasets and performance.\label{tab:1}}
%\smallskip
%\begin{tabular}{cccc}
%\hline
%\textbf{ Offset $\delta$, nm} & \textbf{Image pairs} & \textbf{Localization precision, nm} & \textbf{Training localization precision, nm} \\
%\hline
 %250 & 33562 & $48.4 \pm 0.7$ & $44.2 \pm 0.7$ \\ 
 %500 & 34842 & $40.0 \pm 0.5$ & $35.7 \pm 0.5$ \\ 
 %750 & 21685 & $44.0 \pm 0.7$ & $40.6 \pm 0.7$ \\ 
%\hline
%\end{tabular}
%\end{table}

%\begin{figure}[htbp]
%\centering
%\includegraphics[width=.4\textwidth]{example-image-a}
%\qquad
%\includegraphics[width=.4\textwidth]{example-image-b}
%\caption{Always give a caption.\label{fig:i}}
%\end{figure}

\begin{figure*}[ht]
\centering
\subfloat[][\label{fig:3_d500}${\Delta{z}_{k}}^{pred}$ vs ${\Delta{z}_{k}}^{test}$ for $\delta = 500$~nm]
{\includegraphics[ width=.48\textwidth]{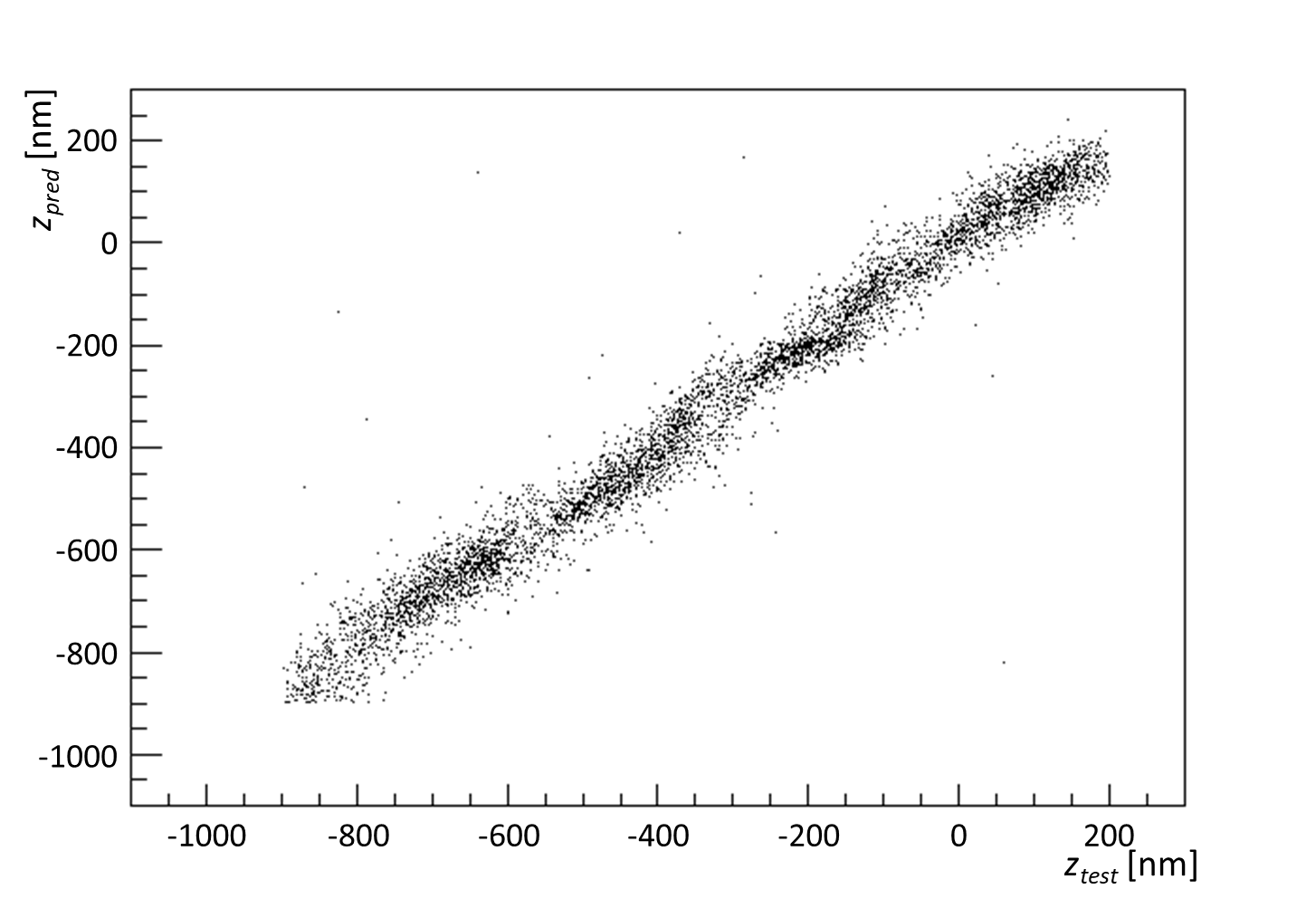}}
\subfloat[][\label{fig:4_d500}Prediction precision for $\delta = 500$~nm]
{\includegraphics[ width=.48\textwidth]{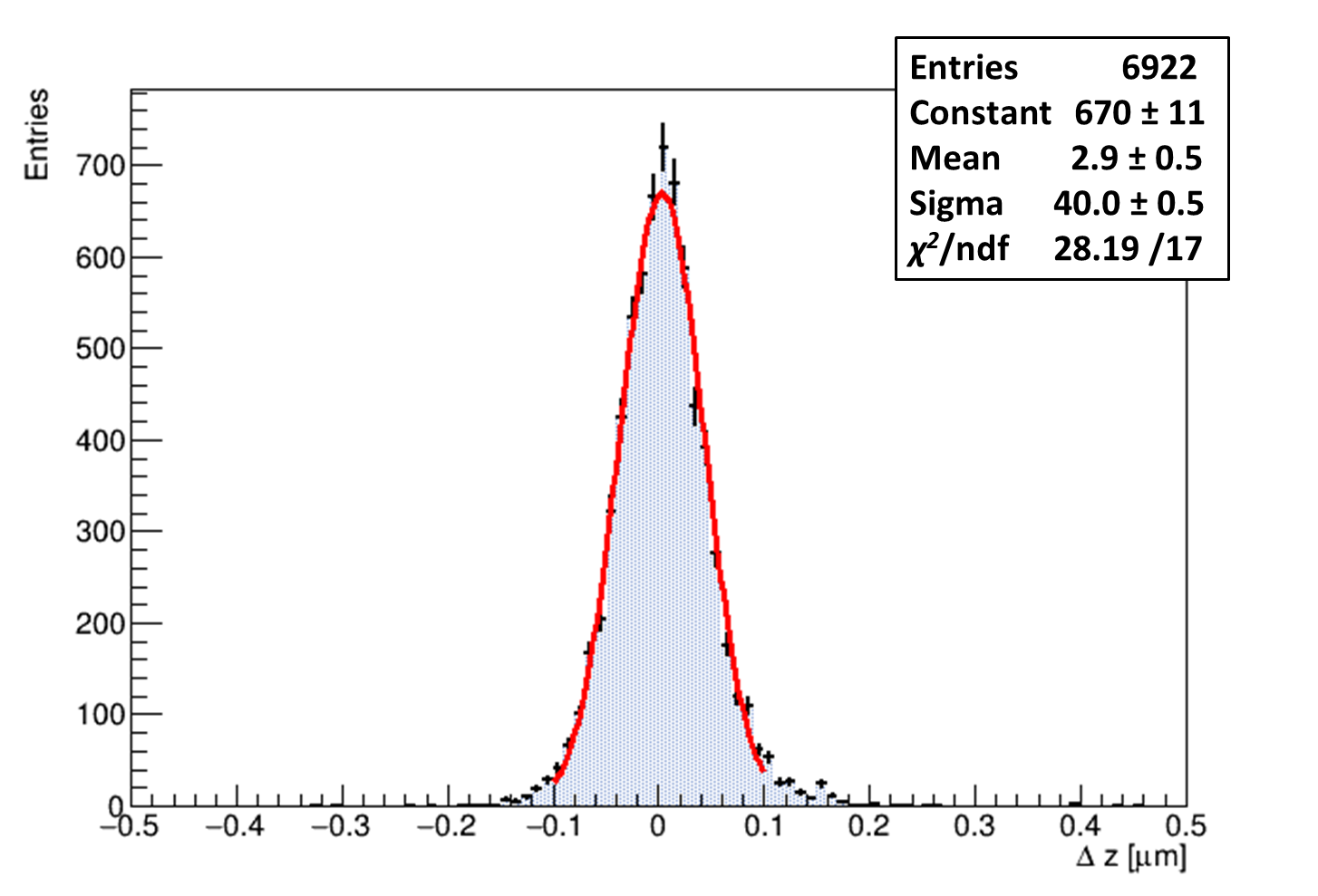}}\\
\subfloat[][\label{fig:4_d250}Prediction precision for $\delta = 250$~nm]
{\includegraphics[ width=.48\textwidth]{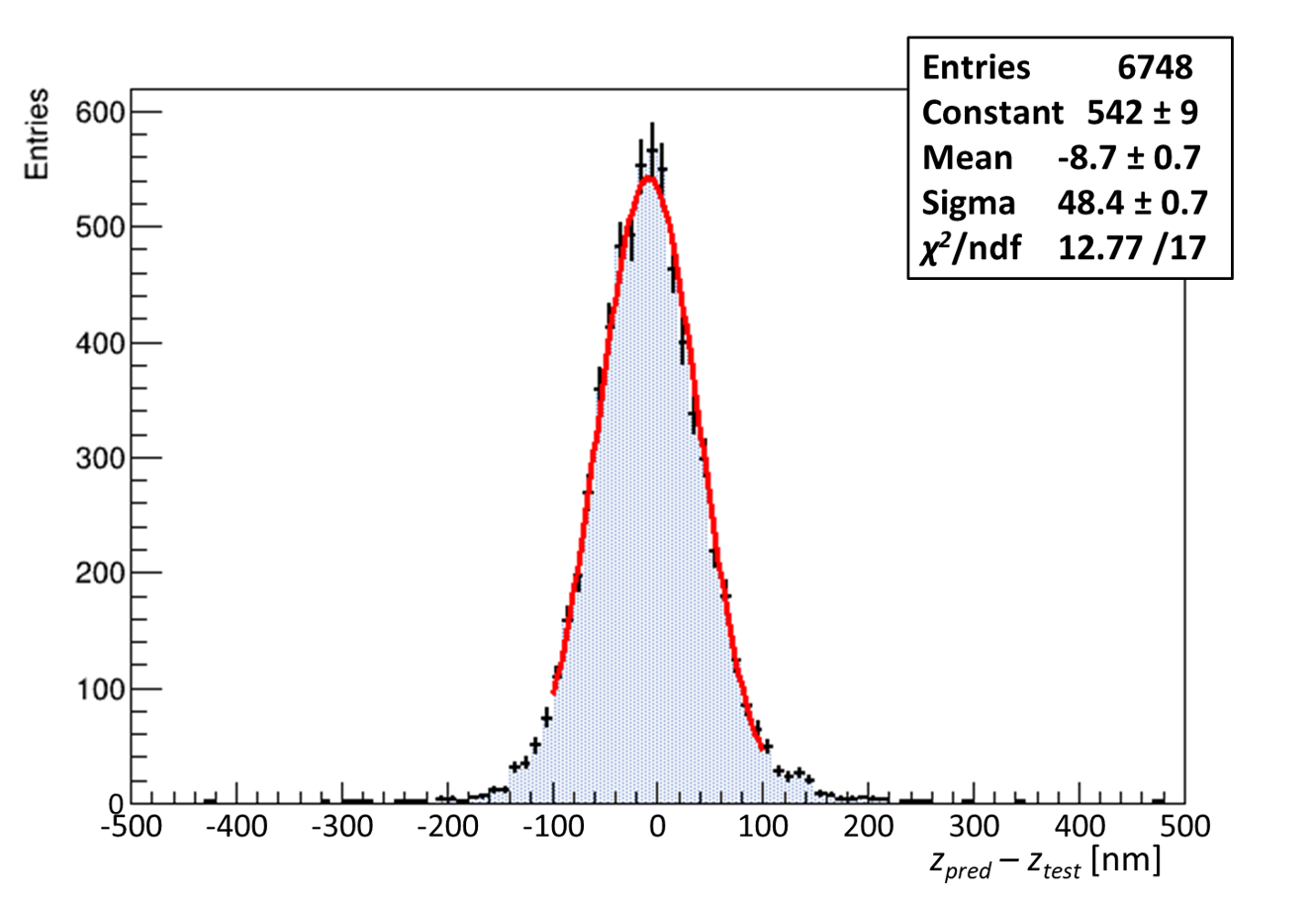}}
\subfloat[][\label{fig:4_d750}Prediction precision for $\delta = 750$~nm]
{\includegraphics[ width=.48\textwidth]{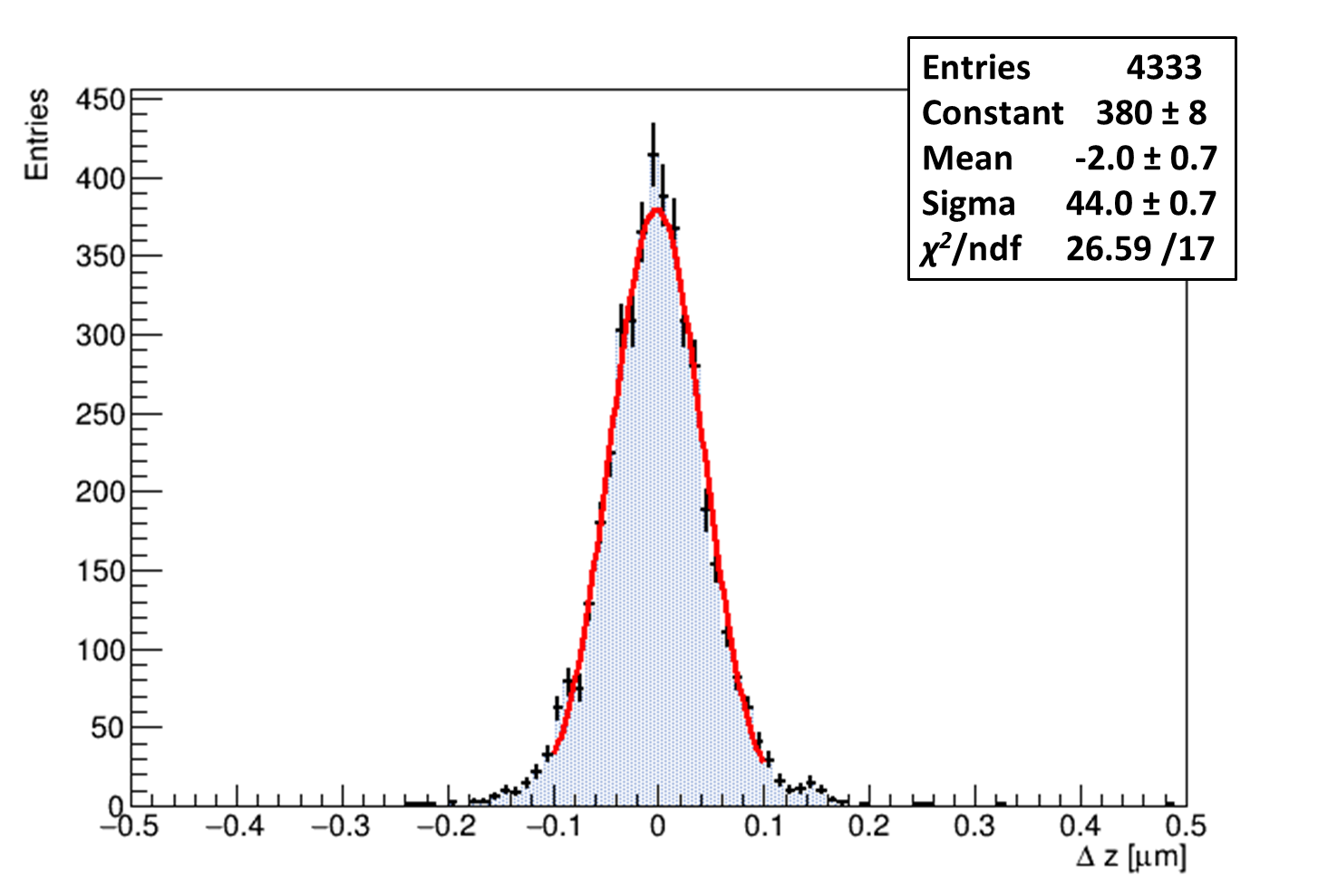}}
\caption{(a) Predicted value ${\Delta{z}_{k}}^{pred}$ as a function of test value ${\Delta{z}_{k}}^{test}$ for image offset $\delta = 500$~nm. Prediction precision $({\Delta{z}_{k}}^{pred} - {\Delta{z}_{k}}^{test})$ for image offsets $\delta$ equal to (b) 500~nm, (c) 250~nm and (d) 750~nm.
} \label{fig:cnn_perf}
\end{figure*}

\subsection{Measurement results and performance}
\label{sec:train_res}

After training, CNN performance was tested on a fraction of the image dataset that was not used for training ($\sim{20}\%$ of the entire dataset). The plot in Figure~\ref{fig:3_d500} shows the proportionality of the predicted defocus values to the expected ones, as well as the independence of performance from the defocus value for image offset $\delta=500$~nm. For image offsets $\delta=250$~nm and $\delta=750$~nm the corresponding plots are similar. The periodic structure in the proportionality plot that repeats every $250$~nm is due to the image acquisition scheme in which the images are taken with the same sampling step in the range $245-255$~nm.

The difference between the predicted defocus values ${\Delta{z}_{k}}^{pred}$ and the calculated ones ${\Delta{z}_{k}}^{test}$ defines the precision of the CNN and is shown in Figure~\ref{fig:4_d500}, ~\ref{fig:4_d250} and Figure~\ref{fig:4_d750} for image offsets $\delta=500$~nm,  $\delta=250$~nm and  $\delta=750$~nm, respectively. The axial localization precision $\sigma_{z}$ is defined by the width of the difference distribution $({\Delta{z}_{k}}^{pred} - {\Delta{z}_{k}}^{test})$ and is summarized in the third column of Table~\ref{tab:1} for each image dataset. These values, obtained from the test set, are compatible with the localization precision achieved on the training data, shown in the fourth column of Table~\ref{tab:1}. The training precision values are slightly better than the test ones, which is expected due to the reduced variability in the training dataset and the optimization bias introduced during model fitting. In ideal conditions, where the model generalizes well and overfitting is absent, the residual distributions for training and test sets are expected to be similar. However, the test set may contain more diverse examples and is not directly optimized during training, which can lead to a slightly broader residual distribution and thus a marginally higher localization uncertainty.

%The goodness of fit of the Gaussian model applied to the CNN prediction residuals (red solid lines in plots shown in Figure~\ref{fig:cnn_perf}) was evaluated with the reduced chi-squared values ($\chi^2/\mathrm{ndf}$) for each histogram shown in Figure~\ref{fig:cnn_perf}). The results are: $12.77/17$ ($\approx 0.75$) for 250~nm, $28.19/17$ ($\approx 1.66$) for 500~nm, and $26.59/17$ ($\approx 1.56$) for 750~nm. The corresponding p-values are 0.751, 0.0428, and 0.0644, respectively. These values suggest that the simple Gaussian fit is good enough. We have also checked that the use of a gaussian+constant does not improve the goodness of the fit. 

The red dotted line in Figure~\ref{fig:Z_prof} represents the fitting of the brightness profile. 
To model the brightness distribution along the optical axis, we used a Gaussian function to represent the nanoparticle signal and added a constant term to account for the background intensity. This approach, often referred to as a Gaussian fit with offset, allows for more accurate estimation of the peak position by separating signal from baseline noise.
The width of the fit allows for a rough estimate of the axial localization precision of the single-focal-plane microscope used for image acquisition to be approximately $\sigma_{z}^{single}\approx{260}$~nm.

In case when two images are used in combination with the reported approach, the axial localization precision improves significantly and reaches $\sigma_{z}^{\delta=500}=40$~nm for the offset $\delta=500$~nm. Somewhat lower performance for offsets $\delta=250$~nm and $\delta=750$~nm implies that offset $\delta=500$~nm is closer to the optimal value, whose existence is demonstrated in ref.~\cite{MUM3}. To put it simply, images with a distance of $\delta=250$~nm are too close and their differences are insufficient to provide the best localization precision. In contrast, images with a distance of $\delta=750$~nm are too far apart. In the latter case, the brightness of one of the images is often too weak, leading to the loss of details due to a too low signal-to-noise ratio.

%The optical resolution may be characterized as $2.9\sigma_{z}$, which is derived from fitting a Gaussian to the point spread function (PSF). 
%The PSF is approximately the same as the brightness profile, given that the diameter of the nanoparticle is significantly smaller than the wavelength of the light used for illumination.
%Therefore, the longitudinal optical resolution of the microscope in use is $R_{z}\approx{750}$~nm.
%When only one image is available to measure the axial coordinate of a nanoparticle, which is often the case in the real-time monitoring of a moving nanoparticle, it is natural to assign the current Z coordinate reading from the microscope.
%The axial location precision in this case is determined by the width of the brightness profile and is approximately $\sigma_{z}^{single}\approx{260}$~nm.

\section{Discussion}

The proposed approach enables the instantaneous measurement of the axial coordinate of a nanoparticle without requiring refocusing or capturing a stack of images at different axial coordinates. The microscope setup requires at least two focal planes offset by a few hundred nanometers. This approach can be easily extended to calculate all three spatial coordinates or to incorporate more than two focal planes, enhancing its flexibility. Moreover, it does not involve explicit fitting to point spread function (PSF) models, making it applicable to complex objects with nontrivial optical properties.

For this study, we utilized an image dataset obtained with a single-focal-plane microscope by sequential image acquisition at the axial step in the range $245-255$~nm. To simulate a dual-focal-plane microscope, we selected image pairs spaced by specific distances: 250, 500, or 750~nm. The nanoparticles were immobilized in gelatin, ensuring negligible displacement during image acquisition. However, mechanical motion during image capture introduced small variations in focal plane distance, which impacted the precision of CNN predictions. Despite this, the achieved axial localization precision of 40~nm, as shown in Figure~\ref{fig:4_d500}, represents a six-fold improvement over traditional single-focal-plane microscopes (see Figure~\ref{fig:Z_prof}).

The reported deep learning approach can be integrated with the existing 2D super-resolution technique~\cite{alexandrov2023super} to extend resolution along the axial direction, leading to a full 3D super-resolution imaging system. This advancement would enable the precise reconstruction of low-energy, sub-micron ion tracks in nano-grained nuclear emulsions, improving the sensitivity of dark matter detectors and addressing challenges related to the neutrino floor background~\cite{NEWSdm:2017efa}. Additionally, in the field of nuclear physics, it will facilitate accurate studies of proton-nucleus fragmentation, supporting applications in hadron therapy for cancer treatment and radiation protection in space environments by providing direct measurements of short-range nuclear fragments~\cite{DAMON2024}.

This study highlights the capability of deep learning to surpass conventional limitations in axial localization, achieving nanometric precision with broad multidisciplinary implications. While the immediate applications include dark matter detection, proton therapy, and space radiation protection, the methodology is highly generalizable. By providing a robust, transparent, and scalable framework, this approach paves the way for future innovations in super-resolution imaging, knowledge extraction, and real-time nanometric tracking across multiple scientific and technological domains.

\appendix
\section{Multifocal plane optical microscopy}

Conventional wide-field optical microscopes have limited axial discrimination, making it difficult to accurately determine the axial coordinate of objects, especially near the focal plane. Even with high numerical aperture objectives, a point source displaced several hundred nanometers from focus often produces nearly identical images, complicating axial localization. As shown in ref.~\cite{MUM1}, localization precision depends strongly on defocus: while in-focus imaging provides good precision in the transverse plane, it performs poorly along the axial axis; high defocus improves axial precision but degrades transverse precision. This trade-off hinders uniform 3D localization.
To address this issue, the multifocal plane imaging technique was proposed in ref.~\cite{MUM2}. 
This technique simultaneously captures two or more separate focal planes within the specimen.
The dual-focal-plane imaging technique is a specific case in which one focal plane matches the typical focal plane imaged by a standard wide-field microscope, whereas the other plane is offset from the standard focal plane. When the object of interest is imaged in the described setup, the image of the offset focal plane offers extra details about the object's axial coordinate. Utilization of this supplementary information enables a more precise determination of the object's 3D location.

\section{Experimental platform}

The hardware used in this study is a desktop computer with 128 GB RAM, 20 core, 3.40 GHz frequency and equipped with an Unix-based operating system. The CNN used in this study was implemented using the Python programming language and, in particular, the modules Keras \cite{keras_lib} and Tensorflow \cite{tensorflow_lib}.

\section{Optical microscope setup}

The microscope~\cite{alexandrov2020super} is equipped with a high magnification objective lens with high numerical aperture (Nikon CFI Plan Apo Lambda 100$\times$/1.45 Oil). Placing an extra magnification lens in front of the camera results in an overall magnification of 260$\times$. The critical-type custom illumination system incorporates the blue LED light source (Luminus CBT-120) with a wavelength of ($460 \pm 25)$~nm. The microscope is set up to function in reflection mode, which results in an enhanced signal-to-noise ratio. 
The system is equipped a fast 4-megapixel monochrome camera (Allied Vision Technologies Bonito CL-400B) operating at 100 fps. The sample is positioned within the horizontal plane using a motorized stage (Micos MS-8), and the axial adjustment is accomplished by shifting the objective lens and the entire optical arrangement with a linear stage (Micos UPM-160). The microscope is also fitted with pneumatic vibration dampers (Fabreeka PLM 1).
A Dell Precision T7500 workstation manages the microscope elements, featuring a Matrox Radient eCL framegrabber, a National Instruments PCI-7344 motion control board, and a GeForce GTX 780 GPU for enhanced image processing speed. The LASSO software framework~\cite{LASSO_link,LASSO_article,LASSO_NewHW} offers modules designed for real-time control of microscopes, featuring completely automated image capture and data analysis functionalities.

\acknowledgments

This work has been partially supported by Spoke 1 "FutureHPC \& BigData" of ICSC - Centro Nazionale di Ricerca in High-Performance-Computing, Big Data and Quantum Computing, funded by European Union - NextGenerationEU.

The authors would like to acknowledge the support of the Russian Science Foundation under the program "Conducting Basic Scientific Research and Search Scientific Research by Selected Scientific Groups" (Project No. 23-12-00054).

% Bibliography

%% [A] Recommended: using JHEP.bst file
\bibliographystyle{JHEP}
\bibliography{main}

%% or
%% [B] Manual formatting (see below)
%% (i) We suggest to always provide author, title and journal data or doi:
%% in short all the informations that clearly identify a document.
%% (ii) please avoid comments such as "For a review'', "For some examples",
%% "and references therein" or move them in the text. In general, please leave only references in the bibliography and move all
%% accessory text in footnotes.
%% (iii) Also, please have only one work for each \bibitem.

%\begin{thebibliography}{99}

%\bibitem{a}
%Author,
%\emph{Title},
%\emph{J. Abbrev.} {\bf vol} (year) pg.

%\bibitem{b}
%Author,
%\emph{Title},
%arxiv:1234.5678.

%\bibitem{c}
%Author,
%\emph{Title},
%Publisher (year).

%\end{thebibliography}
\end{document}